\documentclass[10pt,conference]{IEEEtran}

\usepackage{amsmath}
\usepackage{amssymb}
\usepackage{emp}

\usepackage{graphicx}

\newcommand{\twobibs}[2]{#2} 

\newcommand{\thesisonly}[1]{}

\newcommand{\bbcaption}[3]{\caption{#2}}

\newtheorem{theo}{\indent Theorem}
\newtheorem{lemm}[theo]{\indent Lemma}
\newtheorem{coro}[theo]{\indent Corollary}

\newcommand{\row}{\mathrm{row}}
\newcommand{\col}{\mathrm{col}}
\newcommand{\Gr}{G_\row}
\newcommand{\Gc}{G_\col}
\newcommand{\Er}{E_\row}
\newcommand{\Ec}{E_\col}

\newcommand{\capacity}{\mathsf{cap}}
\newcommand{\size}[1]{\left|#1\right|}
\newcommand{\set}[1]{\left\{#1\right\}}
\newcommand{\sett}[1]{\{#1\}}

\newcommand{\Ais}{\mathsf{A}} 
\newcommand{\Bis}{\mathsf{B}} 
\newcommand{\Cis}{\mathsf{C}} 
\newcommand{\Dis}{\mathsf{D}} 
\newcommand{\Tis}{\mathsf{T}} 
\newcommand{\Uis}{\mathsf{U}}
\newcommand{\Vis}{\mathsf{V}}

\newcommand{\pis}{\Lambda} 
\newcommand{\integers}{\mathbb{Z}}
\newcommand{\calS}{\mathbb{S}}
\newcommand{\scriptS}{\mathcal{S}}
\newcommand{\scriptF}{\mathcal{F}}

\newcommand{\Prob}{\mathrm{Prob}}
\newcommand{\bfp}{\boldsymbol{p}}
\newcommand{\bfg}{\boldsymbol{g}}
\newcommand{\bfw}{\mathbf{w}}

\newcommand{\interior}{\bar\partial}
\newcommand{\boundary}{\partial}

\newcommand{\lexprec}{\prec_\mathrm{lex}}
\newcommand{\tvprec}{\prec_\mathrm{irs}}
\newcommand{\lexf}{f_\mathrm{lex}}
\newcommand{\tvf}{f_\mathrm{irs}}
\newcommand{\skipprec}{\prec_\mathrm{skip}}

\newcommand{\Ygk}{Y_\gamma^{\langle k \rangle}}
\newcommand{\Zgk}{Z_\gamma^{\langle k \rangle}}
\newcommand{\gk}{_\gamma^{\langle k \rangle}}
\newcommand{\rhok}{\rho^{\langle k \rangle}}
\newcommand{\ck}{c^{\langle k \rangle}}
\newcommand{\ygk}{y}
\newcommand{\zgk}{z}
\newcommand{\pygk}{p_{\gamma,\ygk}^{\langle k \rangle}}
\newcommand{\pzgk}{p_{\gamma,\zgk}^{\langle k \rangle}}
\newcommand{\barpygk}{\bar{p}_{\gamma,\ygk}^{\langle k \rangle}}
\newcommand{\barpzgk}{\bar{p}_{\gamma,\zgk}^{\langle k \rangle}}
\newcommand{\Upsgk}{\Upsilon\gk}
\newcommand{\Psigk}{\Psi\gk}
\newcommand{\striplength}{\theta}
\newcommand{\mycomplement}{\mathrm{comp}}
\newcommand{\Wnew}{W^\mathrm{new}}
\newcommand{\ncapacity}{\widehat{\mathsf{cap}}}
\newcommand{\sfm}{\mathsf{m}}

\newcommand{\Footnotemark}[1]{${}^{#1}$}
\newcommand{\Footnotetext}[2]{\begin{figure}[b]\footnotesize%
\vspace{-3ex}\hrulefill\hfill\hfill\makebox[0em]{}%
\vfill
\par${}^{#1}$ #2\vspace{-0.60ex}\end{figure}\addtocounter{figure}{0}}

\begin{document}

\title{Concave Programming Upper Bounds on the Capacity of 2-D Constraints\Footnotemark{*}}

\author{\authorblockN{Ido Tal \qquad Ron M. Roth}
\authorblockA{
Computer Science Department,\\
Technion, Haifa 32000, Israel.\\
Email: {\tt idotal@ieee.org, ronny@cs.technion.ac.il}
}}
\maketitle

\begin{abstract}
The capacity of 1-D constraints is given by the entropy of a corresponding stationary maxentropic Markov chain. Namely, the entropy is maximized over a set of probability distributions, which is defined by some linear requirements. In this paper, certain aspects of this characterization are extended to 2-D constraints. The result is a method for calculating an upper bound on the capacity of 2-D constraints.

The key steps are: The maxentropic stationary probability distribution on square configurations is considered. A set of linear equalities and inequalities is derived from this stationarity. The result is a concave program, which can be easily solved numerically. Our method improves upon previous upper bounds for the capacity of the 2-D ``no independent bits'' constraint, as well as certain 2-D RLL constraints.
\end{abstract}
\Footnotetext{*}{This work was supported by grant
       No.~2002197 from
       the United-States--Israel Binational Science Foundation (BSF),
       Jerusalem, Israel.}
%
%
\section{Introduction}%
Let $\Sigma$ be a finite alphabet. A one-dimensional (1-D) constraint is a set $S$ of words over $\Sigma$. For the set $S$ to be called a 1-D constraint, there must exist an edge-labeled graph $G$ with the following property: a word $\bfw = w_1 w_2 \ldots w_n$ is in $S$ iff there exists a path in $G$ for which the successive edge labels are $w_1,w_2,\ldots,w_n$ (see~\cite{MarcusRothSiegel:98}).

A two dimensional (2-D) constraint over $\Sigma$ is a generalization of a 1-D constraint; it is a set $\calS$ of rectangular configurations over $\Sigma$ and is defined through a pair of \emph{vertex}-labeled graphs $(\Gr,\Gc)$, where $\Gr = (V,\Er,L)$ and $\Gc=(V,\Ec,L)$. Namely, both graphs share the same vertex set and the same vertex labeling function $L : V \to \Sigma$. The constraint $\calS=\calS(\Gr,\Gc)$ consists of all finite rectangular configurations $(w_{i,j})$ over $\Sigma$ with the following property: Let $\Ais$ be the rectangular index set of $(w_{i,j})_{(i,j) \in \Ais}$. There exists a configuration $(u_{i,j})_{(i,j) \in \Ais}$ over the vertex set $V$ such that (a) for each $(i,j)\in \Ais$ we have $w_{i,j} = L(u_{i,j})$; (b) each row in $(u_{i,j})$ is a path in $\Gr$; (c) each column in $(u_{i,j})$ is a path in $\Gc$. Examples of 2-D constraints include the square constraint~\cite{WeeksBlahut:98}, 2-D runlength-limited (RLL) constraints~\cite{KatoZeger:99},
2-D symmetric runlength-limited (SRLL) constraints~\cite{Etzion:97}, and the ``no isolated bits'' constraint \cite{Halevy+:04}.

Let $\calS$ be a given 2-D constraint over a finite alphabet $\Sigma$. Denote
by $\Sigma^{M \times N}$ the set of $M \times N$ configurations over $\Sigma$, and let
\[
\calS_{M,N} = \calS \cap \Sigma^{M \times N} \; , \quad \calS_M = \calS \cap \Sigma^{M \times M} \; .
\]
The capacity of $\calS$ is equal to
\begin{equation}
\label{eq:KZCapacity}
\capacity(\calS) = \lim_{M \to \infty}
\frac{1}{M^2} \cdot \log_2 \size{\calS_M} \; .
\end{equation}

In this paper, we show a method for calculating an upper bound on
$\capacity(\calS)$. Two other methods of calculating an upper bound on the capacity of a 2-D constraint are the following: The first method is the so called ``stripe method,'' in which we fix a positive integer $N$, and bound $\capacity(\calS)$ by
\begin{equation}
\label{eq:stripeMethod}
\capacity(\calS) \leq \lim_{M \to \infty} \frac{1}{M \cdot N} \cdot \log_2 \size{\calS_{M,N}} \; .
\end{equation}
Namely, we consider only stripes of width $N$, and essentially get a 1-D constraint (since we may regard each of the possible row values as a letter in an auxiliary alphabet). The RHS of (\ref{eq:stripeMethod}) is easily calculated for modest values of $N$: Let $G$ be the edge-labeled graph corresponding to the 1-D constraint, and let $A_G$ be the adjacency matrix of $G$. Denote by $\lambda(A_G)$ the Perron eigenvalue of $A_G$. By \cite[\S 3.2]{MarcusRothSiegel:98}, the RHS of (\ref{eq:stripeMethod}) is equal to $\lambda(A_G)$. The second method for upper-bounding $\capacity(\calS)$ is the generalization presented by Forchhammer and Justesen \cite{ForchhammerJustesen:00} to the method of Calkin and Wilf \cite{CalkinWilf:97}.

The capacity of a given 1-D constraint is known to be equal to the value of an optimization program, where the optimization is on the entropy of a certain stationary Markov chain, and is carried out over the conditional probabilities of that chain (see~\cite[\S 3.2.3]{MarcusRothSiegel:98}). We try to extend certain aspects of this characterization of capacity to 2-D constraints. What results is a (generally non-tight) upper bound on $\capacity(\calS)$.

The structure of this paper is as follows.
In Section~\ref{sec:convex:notation}, we set up some notation. Then, in Section~\ref{sec:stationarity}, we show the existence of a certain stationary random variable taking values on $\calS_M$ and having entropy approaching the capacity of $\calS$, as $M \to \infty$. We then consider a relatively small sub-configuration of that random variable, and denote it by $X^{(M)}$. The section concludes with an upper bound on the capacity of $\calS$, which is a function of the probability distribution of $X^{(M)}$. In Section~\ref{sec:linearEquations}, we derive a set of linear equations which hold on the probability distribution of $X^{(M)}$. In Section~\ref{sec:upperBound}, we argue as follows: The bound derived in Section~\ref{sec:stationarity} is a function of the probability distribution of $X^{(M)}$, which we do not know how to calculate; however, by Section~\ref{sec:linearEquations} we know that this probability distribution is subject to a set of linear requirements. Thus, we formalize an optimization problem, where the unknown probability distribution is
replaced by a set of variables, subject to the above-mentioned linear requirements. The maximum of this optimization problem is an upper bound on the capacity of $\calS$. We then show that this optimization problem is easily solved, since it is an instance of convex programming. In Section~\ref{sec:compres}, we
show our computational results. Finally, in Section~\ref{sec:asymptotics} we present an asymptotic analysis of our method.

We note at this point that although this paper
deals with 2\hbox{-}D constraints, our method can be easily generalized to higher dimensions as well.

\section{Notation}%
\label{sec:convex:notation}
This section is devoted to setting up some notation.

\subsection{Index sets and configurations}
Denote the set of integers by $\integers$. A (2-D) index set $\Uis \subseteq \integers^2$ is a set of integer pairs. A 2-D configuration over $\Sigma$ with an index set $\Uis$ is a function $w : \Uis \to \Sigma$. We denote such a configuration as $w = (w_{i,j})_{(i,j) \in \Uis}$, where for all $(i,j) \in \Uis$, we have that $w_{i,j} \in \Sigma$. In this paper, index sets will always be denoted by upper-case Greek letters or upper-case Roman letters in the sans-serif font. Since many of our configurations will be $M \times N$, we have set aside special notation for their index sets; let
\[
\Bis_{M,N} = \set{(i,j) : 0 \leq i < M \; , \quad 0 \leq j < N } \; .
\]
Also, denote
\[
\Bis_{M} = \Bis_{M,M} = \set{(i,j) : 0 \leq i,j < M } \; .
\]

For integers $\alpha, \beta$ we denote the shifting of $\Uis$ by $(\alpha,\beta)$ as
\[
\sigma_{\alpha,\beta}(\Uis) = \set{(i+\alpha,j+\beta) : (i,j) \in \Uis} \; .
\]
Moreover, by abuse of notation,
let $\sigma_{\alpha,\beta}(w)$ be the shifted configuration (with
index set $\sigma(\Uis)$):
\[
\sigma_{\alpha,\beta}(w)_{i+\alpha, j + \beta} = w_{i,j} \; .
\]

For a configuration $w$ with index set $\Uis$, and an index set $\Vis \subseteq \Uis$, denote the restriction of $w$ to $\Vis$ by $w[\Vis] = (w[\Vis]_{i,j})_{(i,j) \in \Vis}$; namely,
\[
w[\Vis]_{i,j} = w_{i,j} \; , \quad \mbox{where} \quad (i,j) \in \Vis \; .
\]

We denote the restriction of $\calS$ to $\Uis$ by $\calS[\Uis]$:
\begin{equation}
\label{eq:restriction}
\calS[\Uis] = \set{ w : \mbox{there exists $w' \in \calS$ such that $w'[\Uis] = w$}} \; .
\end{equation}

\subsection{Strict total order}

A \emph{strict total order} $\prec$ is a relation on $\integers^2 \times \integers^2$, satisfying the following conditions for all $(i_1,j_1), (i_2,j_2),(i_3,j_3) \in \integers^2$.
\begin{itemize}
\item If $(i_1,j_1) \neq (i_2,j_2)$, then either $(i_1,j_1) \prec (i_2,j_2)$ or $(i_2,j_2) \prec (i_1,j_1)$, but not both.
\item If $(i_1,j_1) = (i_2,j_2)$, then neither $(i_1,j_1) \prec (i_2,j_2)$ nor $(i_2,j_2) \prec (i_1,j_1)$.
\item If $(i_1,j_1) \prec (i_2,j_2)$ and $(i_2,j_2) \prec (i_3,j_3)$, then $(i_1,j_1) \prec (i_3,j_3)$.
\end{itemize}
For $(i,j) \in \integers^2$, define $\Tis_{i,j}^{(\prec)}$ as all the indexes preceding $(i,j)$. Namely,
\[
\Tis_{i,j}^{(\prec)} = \set{(i',j') \in \integers^2: (i',j') \prec (i,j)} \; .
\]

\subsection{Entropy}
Let $X$ and $Y$ be two random variables. Denote
\[
p_x = \Prob(X = x) \; .
\]
and
\[
\quad p_{y|x} = \Prob(X = x, Y = y)/\Prob(X=x) \; .
\]
The entropy of $X$ is denoted by $H(X)$ and is equal to
\[
H(X) = \sum_{x} p_x \log p_x \; ,
\]
where the sum is on all $x$ for which $\Prob(X = x)$ is positive. Similarly, we define the conditional entropy $H(Y|X)$ as
\[
H(Y|X) = \sum_x p_x \sum_y p_{y|x} \log p_{y|x} \; ,
\]
where we sum on all $x$ for which $p_x$ is positive and all $y$ for which $p_{y|x}$ is positive.

\section{A preliminary upper bound on $\capacity(\calS)$}
\label{sec:stationarity}
Let $M$ be a positive integer and let $W$ be a random variable taking values on $\calS_M$. We say that $W$ is \emph{stationary} if for all $\Uis \subseteq \Bis_M$, all $\alpha,\beta \in \integers$ such that $\sigma_{\alpha,\beta}(\Uis) \subseteq \Bis_M$, and all $w' \in \calS[\Uis]$, we have that
\[
\Prob( W[\Uis] = w' ) = \Prob( W[\sigma_{\alpha,\beta}(\Uis)] = \sigma_{\alpha,\beta}(w')) \; .
\]

The following is a corollary
of \cite[Theorem 1.4]{BurtonSteif:94}. The proof is given in the Appendix.
\begin{theo}
\label{theo:burtonSteif}
There exists a series of random variables $(W^{(M)})_{M=1}^\infty$ with the following properties:
(i) Each $W^{(M)}$ takes values on $\calS_M$.
(ii) The probability distribution of $W^{(M)}$ is stationary.
(iii) The normalized entropy of $W^{(M)}$ approaches $\capacity(\calS)$,
\begin{equation}
\label{eq:calSasHW}
\capacity(\calS) = \lim_{M \to \infty}
\frac{1}{M^2} \cdot H(W^{(M)}) \; .
\end{equation}
\end{theo}

We now proceed towards deriving Lemma~\ref{lemm:capacityBoundedByHYZ} below, which gives an upper bound on $\capacity(\calS)$, and makes use of the stationarity property. We note in advance that this bound is not actually meant to be calculated. Thus, its utility will be made clear in the following sections. In order to enhance the exposition, we accompany the derivation with two running examples.

\textbf{Running Example I:} Define the lexicographic order $\lexprec$ as follows:
$(i_1,j_1) \lexprec (i_2,j_2)$ iff
\begin{itemize}
\item $i_1 < i_2$, or
\item ($i_1 = i_2$ and $j_1 < j_2$).
\end{itemize}

\textbf{Running Example II:} Define the ``interleaved raster scan'' order $\tvprec$ as follows:
$(i_1,j_1) \tvprec (i_2,j_2)$ iff
\begin{itemize}
\item $i_1 \equiv 0 \pmod{2}$ and  $i_2 \equiv 1 \pmod{2}$, or
\item $i_1 \equiv i_2 \pmod{2}$ and $i_1 < i_2$, or
\item $i_1 = i_2$ and $j_1 < j_2$.
\end{itemize}
(See Figure~\ref{fig:lextvexample} for both examples.)

\begin{figure}
\[
\begin{array}{c}
\begin{array}{|c|c|c|c|c|}
\hline
1 & 2 & 3 & 4 & 5 \\ \hline
6 & 7 & 8 & 9 & 10 \\ \hline
11 & 12 & 13 & 14 & 15 \\ \hline
16 & 17 & 18 & 19 & 20 \\ \hline
21 & 22 & 23 & 24 & 25 \\ \hline
\end{array}
\\ \\
\lexprec
\end{array}
\qquad
\begin{array}{c}
\begin{array}{|c|c|c|c|c|}
\hline
1 & 2 & 3 & 4 & 5 \\ \hline
16 & 17 & 18 & 19 & 20 \\ \hline
6 & 7 & 8 & 9 & 10 \\ \hline
21 & 22 & 23 & 24 & 25 \\ \hline
11 & 12 & 13 & 14 & 15 \\ \hline
\end{array}
\\ \\
\tvprec
\end{array}
\]
\bbcaption{הסדרים השלמים $\lexprec$ ו- $\tvprec$.}{An entry labeled $i$ in the left (right) configuration precedes an entry labeled $j$ according to $\lexprec$ ($\tvprec$) iff $i < j$.}{Strict total orders $\lexprec$ and $\tvprec$.}
\label{fig:lextvexample}
\end{figure}

For the rest of this section, fix positive integers $r$ and $s$, and define the index set
\[
\pis = \Bis_{r,s}\; .
\]
We will refer to
$\pis$
as ``the patch.'' The bound we derive in Lemma~\ref{lemm:capacityBoundedByHYZ} will be a function of the following:
\begin{itemize}
\item the strict total order $\prec$,
\item the integers $r$ and $s$, which determine the order $r \times s$ of the patch $\pis$,
\item an integer $c$, which will denote the number of ``colors'' we encounter,
\item a coloring function $f : \integers^2 \to \set{1,2,\ldots,c}$, mapping each point in $\integers^2$ to one of $c$ colors,
\item $c$ indexes, $(a_\gamma,b_\gamma)_{\gamma = 1}^c$, such that for all $1 \leq \gamma \leq c$,
\[
(a_\gamma,b_\gamma) \in \pis
\]
(namely, each color $\gamma$ has a designated point in the patch, \emph{which may or may not be of color $\gamma$}).
\end{itemize}

The function $f$ must satisfy two requirements, which we now elaborate on.
Our first requirement is: for all $1 \leq \gamma \leq c$,
\begin{equation}
\label{eq:asymptoticallyEqualCases}
\lim_{M \to \infty} \frac{\set{(i,j) \in \Bis_M : f(i,j) = \gamma}}{M^2} = \frac{1}{c} \; .
\end{equation}
Namely, as the orders of $W^{(M)}$ tend to infinity, each color is equally\footnote{In fact, it is possible to generalize (\ref{eq:asymptoticallyEqualCases}), and require only that the limit exists for all $\gamma$. We have not found this generalization useful.} likely. Our second requirement is as follows: there exist index sets $\Psi_1,\Psi_2,\ldots,\Psi_c \subseteq \pis$ such that for all indexes $(i,j) \in \integers^2$,
\begin{equation}
\label{eq:PsiInvariable}
\sigma_{i',j'}(\Psi_\gamma) = \Tis_{i,j}^{(\prec)} \cap \sigma_{i',j'}(\pis) \; ,
\end{equation}
where $\gamma = f(i,j)$, $i' = a_\gamma - i$, and $j' = b_\gamma - j$. Namely, let $(i,j)$ be such that $f(i,j) = \gamma$, and shift $\pis$ such that $(a_\gamma,b_\gamma)$ is shifted to $(i,j)$. Now, consider the set of all indexes in the shifted $\pis$ which precede $(i,j)$: this set must be equal to the correspondingly shifted $\Psi_\gamma$.

\begin{figure}[t]
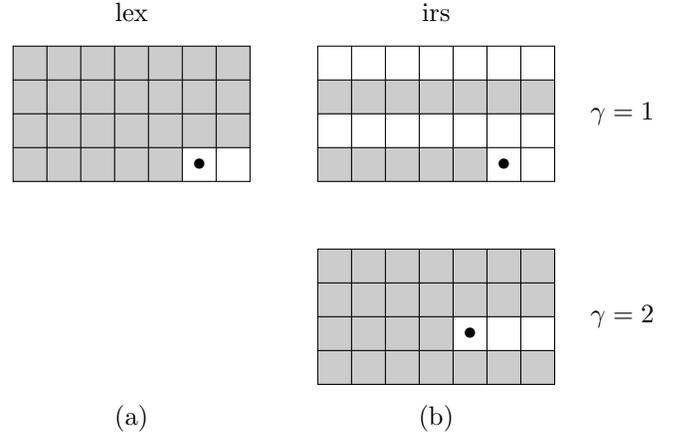

\centering
\begin{empfile}
\begin{emp}(1,1)
numeric u;

u := 0.45cm;

picture grid;
grid := image(
numeric i;
numeric j;

numeric r;
numeric s;

r := 4;
s := 7;

for i=0 upto r:
  draw (0,-u*i)--(u*s,-u*i);
endfor;

for j=0 upto s:
  draw (u*j,0)--(u*j,-u*r);
endfor;
);

picture lexGrid;
lexGrid := image(
numeric a[];
numeric b[];

a1 = 3;
b1 = 5;

fill (0,0)--(u*s,0)--(u*s,-u*(r-1))--(0,-u*(r-1))--cycle withcolor 0.8*white;
fill (0,-u*(r-1))--(u*b1,-u*(r-1))--(u*b1,-u*r)--(0,-u*r)--cycle withcolor 0.8*white;
draw grid;
label(btex $\bullet$ etex, (u*(b1+0.5),-u*(a1+0.5)));
);

draw lexGrid;

picture tvGrid[];
tvGrid1 := image(
numeric a[];
numeric b[];

a1 = 3;
b1 = 5;

fill (0,-u*1)--(u*s,-u*1)--(u*s,-u*2)--(0,-u*2)--cycle withcolor 0.8*white;
fill (0,-u*(r-1))--(u*b1,-u*(r-1))--(u*b1,-u*r)--(0,-u*r)--cycle withcolor 0.8*white;
draw grid;
label(btex $\bullet$ etex, (u*(b1+0.5),-u*(a1+0.5)));
);

draw tvGrid1 shifted (9u,0);

tvGrid2 := image(
numeric a[];
numeric b[];

a2 = 2;
b2 = 4;

fill (0,0)--(u*s,0)--(u*s,-u*(r-2))--(0,-u*(r-2))--cycle withcolor 0.8*white;
fill (0,-u*(r-2))--(u*b2,-u*(r-2))--(u*b2,-u*(r-1))--(0,-u*(r-1))--cycle withcolor 0.8*white;
fill (0,-u*(r-1))--(u*s,-u*(r-1))--(u*s,-u*r)--(0,-u*r)--cycle withcolor 0.8*white;
draw grid;
label(btex $\bullet$ etex, (u*(b2+0.5),-u*(a2+0.5)));
);

draw tvGrid2 shifted (9u,-6u);

label(btex $\textrm{lex}$ etex, (u*3.5,u*1));
label(btex $\textrm{irs}$ etex, (u*12.5,u*1));

label(btex (a) etex, (u*3.5,-u*11));
label(btex (b) etex, (u*12.5,-u*11));

label(btex $\gamma=1$ etex, (u*18,-u*2));
label(btex $\gamma=2$ etex, (u*18,-u*8));
\end{emp}
\end{empfile}

\caption{The left (right) column corresponds to Running Example I (II). The configurations are of order $r \times s$ and represent the index set $\pis$. The $\bullet$ symbol is in position $(a_\gamma,b_\gamma)$. The shaded part is $\Psi_\gamma$.}
\label{fig:PsiRunning}
\end{figure}

\textbf{Running Example I:} Take $r = 4$ and $s = 7$ as the patch orders. Let the number of colors be $c=1$. Thus, we must define $f=\lexf$ as follows: for all $(i,j) \in \integers^2$, $\lexf(i,j) = 1$. Take the point corresponding to the single color as $(a_1=3,b_1=5)$. See also Figure~\ref{fig:PsiRunning}(a).

\textbf{Running Example II:} As in the previous example, take $r = 3$ and $s = 5$ as the patch orders. Let the number of colors be $c=2$. Define $f=\tvf$ as follows:
\[
\tvf(i,j) =
\begin{cases}
1 & i \equiv 0 \pmod{2} \\
2 & i \equiv 1 \pmod{2}
\end{cases} \; .
\]
Take $(a_1=3,b_1=5)$ and $(a_2=2,b_2=4)$. See also Figure~\ref{fig:PsiRunning}(b).

\begin{lemm}
\label{lemm:capacityBoundedByHYZ}
Let $(W^{(M)})_{M=1}^\infty$ be as in Theorem~\ref{theo:burtonSteif} and define
\[
X^{(M)} = W^{(M)}[\pis] \; .
\]
Let $\prec$, $r$, $s$, $c$, $f$, $(\Psi_\gamma)_{\gamma=1}^c$, and $(a_\gamma,b_\gamma)_{\gamma=1}^c$ be given. For $1 \leq \gamma \leq c$, define
\[
\Upsilon_\gamma = \sett{(a_\gamma,b_\gamma)} \cup \Psi_\gamma \; .
\]
Let
\[
Y_\gamma = X^{(M)}[\Upsilon_\gamma] \;\; \mbox{and} \;\;
Z_\gamma = X^{(M)}[\Psi_\gamma]
\]
(note that $Y_\gamma$ and $Z_\gamma$ are functions of $M$). Then,
\[
\capacity(\calS) \leq \limsup_{M \to \infty}  \frac{1}{c} \sum_{\gamma=1}^c H(Y_\gamma|Z_\gamma) \; .
\]
\end{lemm}

\begin{proof}
Let $X$, $W$ and $\Tis_{i,j}$ be shorthand for $X^{(M)}$, $W^{(M)}$ and $\Tis_{i,j}^{(\prec)}$, respectively. First note that
\[
Y_\gamma = W[\Upsilon_\gamma] \;\; \mbox{and} \;\;
Z_\gamma = W[\Psi_\gamma] \; .
\]
We show that
\[
\lim_{M \to \infty} \frac{1}{M^2} H(W) \leq \limsup_{M \to \infty}  \frac{1}{c} \sum_{\gamma=1}^c H(Y_\gamma|Z_\gamma) \; .
\]
Once this is proved, the claim follows from (\ref{eq:calSasHW}).

By the chain rule \cite[Theorem 2.5.1]{CoverThomas:91}, we have
\[
H(W) = \sum_{(i,j) \in \Bis_M} H(W_{i,j} | W[\Tis_{i,j} \cap \Bis_M]) \; .
\]
We now recall (\ref{eq:PsiInvariable}) and define the index set $\interior$ to be the largest subset of $\Bis_M$ for which the following condition holds: for all $(i,j) \in \interior$, we have that
\begin{equation}
\label{eq:interior}
\sigma_{i',j'}(\Psi_\gamma) \subseteq \Bis_M \; ,
\end{equation}
where hereafter in the proof, $\gamma = f(i,j)$, $i' = a_\gamma - i$, and $j' = b_\gamma - j$. Define $\boundary = \Bis_M \setminus \interior$. Note that since $r$ and $s$ are constant, and $\Psi_1,\Psi_2,\ldots,\Psi_c \subseteq \pis$, then
\[
\frac{\size{\boundary}}{M^2} = O(1/M) \; .
\]
Thus, on the one hand, we have
\[
\frac{1}{M^2} \sum_{(i,j) \in \boundary} H(W_{i,j} | W[\Tis_{i,j} \cap \Bis_M]) \leq
\log_2 \size{\Sigma} \cdot O(1/M) \; .
\]
On the other hand, from (\ref{eq:PsiInvariable}) and (\ref{eq:interior}) we have that for all $(i,j) \in \interior$,
\[
\sigma_{i',j'}(\Psi_\gamma) \subseteq \Tis_{i,j} \cap \Bis_M \; .
\]
Hence, since conditioning reduces entropy \cite[Theorem 2.6.5]{CoverThomas:91},
\begin{align*}
& \frac{1}{M^2} \sum_{(i,j) \in \interior} H(W_{i,j} | W[\Tis_{i,j} \cap \Bis_M]) \\
\leq \; & \frac{1}{M^2} \sum_{(i,j) \in \interior} H(W_{i,j} | W[\sigma_{i',j'}(\Psi_\gamma)]) \\
= \; & \frac{1}{M^2} \sum_{(i,j) \in \interior} H(W[\set{(i,j)} \cup \sigma_{i',j'}(\Psi_\gamma)] | W[\sigma_{i',j'}(\Psi_\gamma)]) \\
= \; & \frac{1}{M^2} \sum_{(i,j) \in \interior} H(Y_\gamma | Z_\gamma)
\; ,
\end{align*}
where the last step follows from the stationarity of $W^{(M)}$. Recalling (\ref{eq:asymptoticallyEqualCases}), the proof follows.
\end{proof}

The following is a simple corollary of Lemma~\ref{lemm:capacityBoundedByHYZ}.

\begin{coro}
\label{coro:capacityBoundedByHYZandRho}
Let $(W^{(M)})_{M=1}^\infty$ be as in Theorem~\ref{theo:burtonSteif} and define
\[
X^{(M)} = W^{(M)}[\pis] \; .
\]
Fix positive integers $r$ and $s$. Let $\ell$ be a positive integer, and let $(\rho^{\langle k \rangle})_{k=1}^\ell$ be non-negative reals such that $\sum_{k=1}^\ell \rho^{\langle k \rangle} = 1$. For every $1 \leq k \leq \ell$,
let $\prec^{\langle k \rangle}$, $c^{\langle k \rangle}$, $f^{\langle k \rangle}$, $(\Psi^{\langle k \rangle}_\gamma)_{\gamma=1}^c$, and $(a^{\langle k \rangle}_\gamma,b^{\langle k \rangle}_\gamma)_{\gamma=1}^c$ be given. Also, for $1 \leq \gamma \leq c^{\langle k \rangle}$, let
\[
\Upsgk = \sett{(a\gk,b\gk)} \cup \Psigk \; .
\]
Define
\[
\Ygk = X^{(M)}[\Upsgk] \;\; \mbox{and} \;\;
\Zgk = X^{(M)}[\Psigk]
\]
(note that $\Ygk$ and $\Zgk$ are functions of $M$). Then,
\[
\capacity(\calS) \leq \limsup_{M \to \infty}  \sum_{k=1}^\ell \frac{\rho^{\langle k \rangle}}{c^{\langle k \rangle}} \sum_{\gamma=1}^{c^{\langle k \rangle}} H(\Ygk|\Zgk) \; .
\]
\end{coro}
Corollary~\ref{coro:capacityBoundedByHYZandRho} is the most general way we have found to state our results. This generality will indeed help us later on. However, almost none of the intuition is lost if the reader has in mind the much simpler case of
\begin{multline}
\label{eq:simplestCase}
\ell=1 \;, \quad \rho^{\langle 1 \rangle} = 1 \; , \quad c^{\langle 1 \rangle} = 1 \; , \quad  \prec^{\langle 1 \rangle} = \lexprec \; ,
\\
(a^{\langle 1 \rangle}_1,b^{\langle 1 \rangle}_1)=(r{-}1,t) \; , \quad \mbox{and} \;\;
\Psi_1^{\langle 1 \rangle} = \pis \cap \Tis_{(a^{\langle 1 \rangle}_1,b^{\langle 1 \rangle}_1)} \; ,
\end{multline}
where $0 \leq t < s$. This simpler case was demonstrated in Running Example I.

\section{Linear requirements}
\label{sec:linearEquations}

Recall that $X^{(M)}=W^{(M)}[\pis]$ is an $r \times s$ sub-configuration of $W^{(M)}$, and thus stationary as well. In this section, we formulate a set of linear requirements (equalities and inequalities) on the probability distribution of $X^{(M)}$. For the rest of this section, let $M$ be fixed and let $X$ be shorthand for $X^{(M)}$.

\subsection{Linear requirements from stationarity}

In this subsection, we formulate a set of linear requirements that follow from the stationarity of $X^{(M)}$. Let $x \in \calS[\pis]$ be a realization of $X$. Denote
\[
p_x = \Prob(X = x) \; .
\]
We start with the trivial requirements. Obviously,
we must have for all $x \in \calS[\pis]$ that
\[
p_x \geq 0 \; .
\]
Also,
\[
\sum_{x \in \calS[\pis]} p_x = 1 \; .
\]

Next, we show how we can use stationarity to
get more linear equations on $(p_x)_{x \in \calS[\pis]}$. Let
\[
\pis' = \set{(i,j) : 0 \leq i < r-1\; , \;\; 0 \leq j < s} \; .
\]
For $x' \in \calS[\pis']$ we must have by stationarity that
\begin{equation}
\label{eq:verticalStationarity}
\Prob( X[\pis'] = x' ) = \Prob( X[\sigma_{1,0}(\pis')] = \sigma_{1,0}(x') ) \; .
\end{equation}
As a concrete example, suppose that $r=s=3$. We claim that
\[
\newcommand{\smallstar}{{\textrm{\scriptsize $*$}}}
\newcommand{\zero}{{\textrm{\scriptsize $0$}}}
\newcommand{\one}{{\textrm{\scriptsize $1$}}}
\Prob\left( X =
\renewcommand{\arraystretch}{0.7}
\arraycolsep0.5ex
\begin{array}{ccc}
\one & \zero & \zero \\
\zero & \zero & \one \\
\smallstar & \smallstar & \smallstar
\end{array}
\right)
=
\Prob\left( X =
\renewcommand{\arraystretch}{0.7}
\arraycolsep0.5ex
\begin{array}{ccc}
\smallstar & \smallstar & \smallstar \\
\one & \zero & \zero \\
\zero & \zero & \one
\end{array}\right) \; ,
\]
where $*$ denotes ``don't care''.

Both the left-hand and right-hand sides of (\ref{eq:verticalStationarity}) are
marginalizations  of $(p_x)_x$. Thus, we get a set of linear equations on $(p_x)_x$,
namely, for all $x' \in \calS[\pis']$,
\[
\sum_{x \, : \, x[\pis']=x'} p_x = \sum_{x \, : \, x[\sigma_{1,0}(\pis')]=
\sigma_{1,0}(x')} p_x \; .
\]

To get more equations, we now apply the same rational horizontally, instead of
vertically. Let
\[
\pis'' = \set{(i,j) : 0 \leq i < r\; , \;\; 0 \leq j < s-1} \; .
\]
for all $x'' \in \calS[\pis'']$,
\[
\sum_{x \, : \, x[\pis'']=x''} p_x =
\sum_{x \, : \, x[\sigma_{0,1}(\pis'')]=\sigma_{0,1}(x'')} p_x \; .
\]

\subsection{Linear equations from reflection, transposition, and complementation}
We now show that if $\calS$ is reflection,  transposition, or complementation invariant (defined below), then we can derive yet more linear equations.

Define $v_M(\cdot)$ ($h_M(\cdot)$) as the vertical (horizontal) reflection of a rectangular configuration with $M$ rows (columns). Namely,
\[
(v_M(w))_{i,j} = w_{M-1-i,j} \; , \quad \mbox{and}
\quad (h_M(w))_{i,j} = w_{i,M-1-j} \; .
\]

Define $\tau$ as the transposition of a configuration. Namely,
\[
\tau(w)_{i,j} = w_{j,i} \; .
\]

For $\Sigma = \set{0,1}$, denote by $\mycomplement(w)$ the bitwise complement of a configuration $w$. Namely,
\[
\mycomplement(w)_{i,j} =
\begin{cases}
1 & \mbox{if $w_{i,j} = 0$}\\
0 & \mbox{otherwise} \; .
\end{cases}
\]

We state three similar lemmas, and prove the first. The proof of the other two is similar.
\begin{lemm}
\label{lemm:reflectionInvariant}
Suppose that $\calS$ is such that for all $M > 0$ and $w \in \Sigma^{M \times M}$,
\[
w \in \calS \iff h_M(w) \in \calS \iff v_M(w) \in \calS \; .
\]
Then, w.l.o.g., the probability distribution of $W$ is such that for all $w \in \calS_M$,
\begin{multline} 
\label{eq:reflectionAssumption}
\Prob(W = w ) = \\
\Prob(W = h_M(w)) = \Prob(W = v_M(w)) \; .
\end{multline} 
\end{lemm}

\begin{lemm}
\label{lemm:transposeInvariant}
Suppose that $\calS$ is such that for all $M > 0$ and $w \in \Sigma^{M \times M}$,
\[
w \in \calS \iff \tau(w) \in \calS \; .
\]
Then, w.l.o.g., $W$ is such that for all $w \in \calS_M$,
\begin{equation}
\label{eq:transposeAssumption}
\Prob(W = w ) = \Prob(W = \tau(w)) \; .
\end{equation}
\end{lemm}

\begin{lemm}
\label{lemm:complementInvariant}
Suppose that $\Sigma = \set{0,1}$ and $\calS$ is such that for all $M > 0$ and $w \in \Sigma^{M \times M}$,
\[
w \in \calS \iff \mycomplement(w) \in \calS \; .
\]
Then, w.l.o.g., $W$ is such that for all $w \in \calS_M$,
\begin{equation}
\label{eq:complimentAssumption}
\Prob(W = w ) = \Prob(W = \mycomplement(w)) \; .
\end{equation}
\end{lemm}

\begin{proof}[Proof of Lemma~\ref{lemm:transposeInvariant}]
Let $h$ and $v$ be shorthand for $h_M$ and $v_M$, respectively.
For $M$ fixed, we define a new random variable $\Wnew$ taking values on $\calS_M$, with the following distribution: for all $w \in \calS_M$,
\[
\Prob(\Wnew {=} w ) = \frac{1}{4} \sum_{\substack{w'\in \\ \{w, h(w), v(w), h(v(w))\}}} \Prob(W {=} w' ) \; .
\]
Since $h(h(w)) = v(v(w)) = w$ and $h(v(w)) = v(h(w))$ we get that (\ref{eq:reflectionAssumption}) holds for $\Wnew$. Moreover, by the concavity of the entropy function,
\[
H(W) \leq H(\Wnew) \; .
\]
Thus, the properties defined in Theorem~\ref{theo:burtonSteif} hold for $\Wnew$.
\end{proof}

If the condition of Lemma~$\ref{lemm:reflectionInvariant}$ holds, then we get the following equations by stationarity. For all $x \in \calS[\pis]$,
\[
p_x = p_{v_r(x)} = p_{h_s(x)} \; .
\]

If the condition of Lemma~$\ref{lemm:transposeInvariant}$ holds, then the following holds by stationarity.
Assume w.l.o.g.\ that $r \leq s$, and let
\[
\tilde \pis = \set{(i,j) : 0 \leq i, j < r} \; .
\]
For all $\chi \in \calS[\tilde \pis]$,
\[
\sum_{x \, : \, x[\tilde \pis]=\chi} p_x = \sum_{x \, : \, x[\tilde \pis]=\tau(\chi)} p_x \; .
\]

If the condition of Lemma~$\ref{lemm:complementInvariant}$ holds, then we get the following equations by stationarity. For all $x \in \calS[\pis]$,
\[
p_x = p_{\mycomplement(x)} \; .
\]

\section{An upper bound on $\capacity(\calS)$}
\label{sec:upperBound}
For the rest of this section, let $r$, $s$, $\ell$, $\rho^{\langle k \rangle}$, $\prec^{\langle k \rangle}$, $c^{\langle k \rangle}$, $f^{\langle k \rangle}$, $\Psi^{\langle k \rangle}_\gamma$, and $(a^{\langle k \rangle}_\gamma,b^{\langle k \rangle}_\gamma)$ be given as in Corollary~\ref{coro:capacityBoundedByHYZandRho}. Recall from Corollary~\ref{coro:capacityBoundedByHYZandRho} that we are interested in
$H(Y_\gamma^{\langle k \rangle}|Z_\gamma^{\langle k \rangle})$, in order to bound $\capacity(\calS)$ from above.

As a first step, we fix $M$ and express
$H(Y_\gamma^{\langle k \rangle}|Z_\gamma^{\langle k \rangle})$ in terms of the probabilities $(p_x)_x$ of the random variable $X^{(M)}$. For given $1 \leq k \leq \ell$ and $1 \leq \gamma \leq c^{\langle k \rangle}$, let
\[
\ygk \in \calS[\Upsilon^{\langle k \rangle}_\gamma]
\;\; \mbox{and} \;\;
\zgk \in \calS[\Psi^{\langle k \rangle}_\gamma]
\]
be realizations of $\Ygk$ and
$\Zgk$, respectively. Let
\[
\pygk = \Prob( \Ygk = \ygk ) \quad \mbox{and} \quad \pzgk = \Prob( \Zgk = \zgk )
\]
($\pygk$ and $\pzgk$ are functions of $M$). From here onward, let $p_y$ and $p_z$ be shorthand for $\pygk$ and $\pzgk$, respectively. Both $p_y$ and $p_z$ are marginalizations  of $(p_x)_x$, namely,
\[
p_y = \sum_{x \in \calS[\pis]\, : \,x[\Upsgk] = \ygk} p_x \; , \quad
p_z = \sum_{x \in \calS[\pis]\, : \,x[\Psigk] = \zgk} p_x \; .
\]
Thus, for given $\gamma$ and $k$,
\[
H(\Ygk|\Zgk) = \sum_{y \in \calS[\Upsgk]} -p_y \log_2 p_y +
\sum_{z \in \calS[\Psigk]} p_z \log_2 p_z
\]
is a function of the probabilities $(p_x)_x$ of $X^{(M)}$.

Our next step will be to reason as follows: We have found linear requirements that
the $p_x$'s satisfy and expressed $H(\Ygk|\Zgk)$ as a function of $(p_x)_x$. However,
\emph{we do not know of a way to actually calculate} $(p_x)_x$. So, instead of the
probabilities $(p_x)_x$, consider the \emph{variables} $(\bar{p}_x)_x$. From this line
of thought we get our main theorem.

\begin{theo}
The value of the optimization program given in Figure~\ref{fig:optProgram} is
an upper bound on $\capacity(\calS)$.
\end{theo}

\begin{proof}
First, notice that if we take $\bar{p}_x = p_x$, then (by Section~\ref{sec:linearEquations}) all the requirements which the $\bar{p}_x$'s are
subject to indeed hold, and the objective function is equal to
\[
\sum_{k=1}^\ell \frac{\rho^{\langle k \rangle}}{c^{\langle k \rangle}} \sum_{\gamma=1}^{c^{\langle k \rangle}} H(\Ygk|\Zgk) \; .
\]
So, the maximum is an upper bound on the above equation. Next, by compactness, a maximum indeed exists. Since the maximum is not a function of $M$, the claim now follows from Corollary~\ref{coro:capacityBoundedByHYZandRho}.
\end{proof}


\begin{figure}
\makebox[0ex]{}\hrulefill\makebox[0ex]{}
\vspace{0.8ex}
\\
maximize
\[
\sum_{k=1}^\ell \frac{\rhok}{\ck}
\sum_{\gamma=1}^{\ck}
\Xi(k,\gamma)
\]
over the variables $(\bar{p}_x)_{x \in \calS[\pis]}$, where for
\[
1 \leq k \leq \ell \; , \quad
1 \leq \gamma \leq c^{\langle k \rangle} \; , \quad
y \in \calS[\Upsgk] \; , \quad
z \in \calS[\Psigk] \; ,
\]
we define
\[
\barpygk \triangleq \sum_{x \in \calS[\pis]\, : \,x[\Upsgk] = y} \bar{p}_x \; , \quad
\barpzgk \triangleq \sum_{x \in \calS[\pis]\, : \,x[\Psigk] = z} \bar{p}_x \; ,
\]
\[
\Xi(k,\gamma) \triangleq
\;\; - \!\!\!\!\! \sum_{y \in \calS[\Upsgk]} \barpygk \log_2 \barpygk \;\; + \!\!\!
\sum_{z \in \calS[\Psigk]} \barpzgk \log_2 \barpzgk \; ,
\]
and the variables $\bar{p}_x$ are subject to the following requirements:
\[
\sum_{x \in \calS[\pis]} \bar{p}_x = 1 \; .
\leqno{\mbox{(i)}}
\]
(ii) For all $x \in \calS[\pis]$,
\[
\bar{p}_x \geq 0 \; .
\]
(iii) For all $x' \in \calS[\pis']$,
\[
\sum_{x \, : \, x[\pis']=x'} \bar{p}_x =
\sum_{x \, : \, x[\sigma_{1,0}(\pis')]=\sigma_{1,0}(x')} \bar{p}_x \; .
\]
(iv) For all $x'' \in \calS[\pis'']$,
\[
\sum_{x \, : \, x[\pis'']=x''} \bar{p}_x =
\sum_{x \, : \, x[\sigma_{0,1}(\pis'')]=\sigma_{0,1}(x'')} \bar{p}_x \; .
\]

(v) (If $\calS$ is reflection (resp. complementation) invariant) For all $x \in \calS[\pis]$,
\[
\bar{p}_x = \bar{p}_{v_r(x)} = \bar{p}_{h_s(x)} \quad (\mbox{resp. } \bar{p}_x = \bar{p}_{\mycomplement(x)}) \; .
\]

(vi) (If $\calS$ is transposition invariant)
For all $\chi \in \calS[\tilde \pis]$,
\[
\sum_{x \, : \, x[\tilde \pis]=\chi} \bar{p}_x =
\sum_{x \, : \, x[\tilde \pis]=\tau(\chi)} \bar{p}_x \; .
\]
\vspace{0.8ex}%
\makebox[0ex]{}\hrulefill\makebox[0ex]{}
\caption{Optimization program over the variables $\bar{p}_x$
(assuming w.l.o.g.\ that $r \leq s$). The optimum is an upper bound on
$\capacity(\calS)$.}
\label{fig:optProgram}
\end{figure}

We now proceed to show that the optimization problem in Figure~\ref{fig:optProgram} is an instance of concave programming \cite[p. 137]{BoydVandenberghe:04}, and thus easily calculated. Since the requirements that the variables $(\bar{p}_x)_x$ are subject to are linear, this reduces to showing that the objective function is concave in $(\bar{p}_x)_x$.

\begin{lemm}
The objective function in Figure~\ref{fig:optProgram} is concave in the variables $(\bar{p}_x)_{x \in \calS[\pis]}$, subject to them being non-negative.
\end{lemm}

\begin{proof}
Recall that for all $1 \leq k \leq \ell$ we have that $\frac{\rhok}{\ck}$ is non-negative. Thus, it suffices to prove that for all $1 \leq k \leq \ell$ and $1 \leq \gamma \leq \ck$, the function $\Xi(k,\gamma)$ is concave in the variables $(\bar{p}_x)_x$. So, let $k$ and $\gamma$ be fixed, and let $\bar{p}_y$ and $\bar{p}_z$ be shorthand for $\barpygk$ and $\barpzgk$, respectively.

Recalling the definition of $\barpygk$ and $\barpzgk$ in Figure~\ref{fig:optProgram} and the fact that $\Psigk \subseteq \Upsgk$, we get that
\[
\Xi(k,\gamma) = \sum_{\substack{y \in \calS[\Upsgk]\\z = y[\Psigk]}} -\bar{p}_y \log_2 \frac{\bar{p}_y}{\bar{p}_z} \; .
\]

Thus, it suffices to show that each summand is concave in $(\bar{p}_x)_{x}$. This is indeed the case: let $(\bar{p}_x^{(1)})_{x \in \calS[\pis]}$ and $(\bar{p}_x^{(2)})_{x \in \calS[\pis]}$ be non-negative. Let $0 \leq \xi \leq 1$ be given, and define $(\bar{p}_x^{(3)})_{x \in \calS[\pis]}$ as
\[
\bar{p}_x^{(3)} = \xi \bar{p}_x^{(1)} + (1-\xi) \bar{p}_x^{(2)} \; , \quad x \in \calS[\pis] \; .
\]
For $t = 1,2,3$, denote by $\bar{p}^{(t)}_y$ and $\bar{p}^{(t)}_z$ the marginalizations corresponding to $(\bar{p}_x^{(t)})_x$. Obviously,
\[
\bar{p}_y^{(3)} = \xi \bar{p}_y^{(1)} + (1-\xi) \bar{p}_y^{(2)} \; ,  \quad y \in \calS[\Upsgk]\; .
\]
and
\[
\bar{p}_z^{(3)} = \xi \bar{p}_z^{(1)} + (1-\xi) \bar{p}_z^{(2)} \; , \quad z \in \calS[\Psigk]\; .
\]
We must show that for all $y \in \calS[\Upsgk]$, $z = y[\Psigk]$
\[
\bar{p}^{(3)}_y \log_2 \frac{\bar{p}^{(3)}_y}{\bar{p}^{(3)}_z} \leq
\xi \bar{p}^{(1)}_y \log_2 \frac{\bar{p}^{(1)}_y}{\bar{p}^{(1)}_z} +
(1-\xi)\bar{p}^{(2)}_y \log_2 \frac{\bar{p}^{(2)}_y}{\bar{p}^{(2)}_z} \; .
\]
This is indeed the case, by the log sum inequality \cite[p. 29]{CoverThomas:91}.
\end{proof}

\section{Computational results}
\label{sec:compres}

At this point, we have formulated a concave optimization problem, and wish to
solve it. There are quite a few programs, termed \emph{solvers}, that enable one to do so. Many such solvers --- most of them proprietary --- are hosted on the servers of the NEOS project \cite{Czyzyk+:98}\cite{GroppMore:97}\cite{Dolan+:02}, and the public may submit moderately sized optimization problems to them. We have coded our optimization problems in the AMPL modeling language \cite{Fourer+:02}, and submitted them to NEOS.

Essentially, a solver starts with some initial guess as to the optimizing value of $(\bar{p}_x)_{x \in \calS[\pis]}$, and then iteratively improves the value of the objective function. This process is terminated when the solver decides that it is ``close enough'' to the optimum. Denote by $\widetilde{\bfp} = (\widetilde{p}_x)_{x \in \calS[\pis]}$ this ``close enough'' assignment to the variables. Of course, we must supply an upper bound on $\capacity(\calS)$, not an approximation to one. Thus, let $\widetilde{f}$ and
\[
\widetilde{\bfg} = (\widetilde{g}_x)_x \; , \quad x \in \calS[\pis] \; ,
\]
be the value of the objective function and its gradient at $\widetilde{\bfp}$,
respectively. Obviously, $\widetilde{f}$ is a lower bound on the value of our
optimization problem. For an upper bound\footnote{We remark in passing that if we had chosen to optimize the \emph{dual problem} \cite[p. 215]{BoydVandenberghe:04}, then the ``dual of'' $\widetilde{f}$ would already have been an upper bound. However, we have not managed to state the dual problem in closed form.}, we replace the objective function in
Figure~\ref{fig:optProgram} by
\[
\mathrm{maximize} \quad \left(\widetilde{f} +
\sum_{x \in \calS[\pis]}
\widetilde{g}_x \cdot (\bar{p}_x -\widetilde{p}_x) \right) \; ,
\]
and get a \emph{linear program} (the value of which can be calculated exactly). By concavity, the value of this linear
program is indeed an upper bound. So, we use NEOS yet again to solve it. For the sake of double-checking, we submitted the above optimization problems to two solvers: IPOPT \cite{WachterBiegle:06} and MOSEK.

Before stating our computational results, let us first define one more strict total order, which we have termed the ``skip'' order, $\skipprec$ (see Figure~\ref{fig:skipexample}). We have that $(i_1,j_1) \skipprec (i_2,j_2)$ iff
\begin{itemize}
\item $i_1 < i_2$, or
\item ($i_1 = i_2$ and $j_1 \equiv 0 \pmod{2}$ and $j_2 \equiv 1 \pmod{2}$), or
\item ($i_1 = i_2$ and $j_1 \equiv j_2 \pmod{2}$ and $j_1 < j_2$)
\end{itemize}

\begin{figure}[t]
\[
\begin{array}{c}
\begin{array}{|c|c|c|c|c|c|c|}
\hline
1 & 5 & 2 & 6 & 3 & 7 & 4\\ \hline
8 & 12 & 9 & 13 & 10 & 14 & 11\\ \hline
15 & 19 & 16 & 20 & 17 & 21 & 18\\ \hline
\end{array}
\\ \\
\skipprec
\end{array}
\]
\bbcaption{הסדר השלם $\skipprec$.}{An entry labeled $i$ in the configuration precedes an entry labeled $j$ according to $\skipprec$ iff $i < j$.}{Strict total order $\skipprec$.}
\label{fig:skipexample}
\end{figure}

Our computational results appear in Table~\ref{tbl:ourUpperBounds}. To the best of our knowledge, they are presently the tightest. The penultimate column contains upper bounds obtained by the method described in \cite{ForchhammerJustesen:00}. When available, these compared-to bounds are taken from previously published work, as indicated to the right of them. The rest are the result of our implementation of \cite{ForchhammerJustesen:00}. For reference, the last column contains corresponding lower bounds. We note that the indexes $(a\gk,b\gk)$ and coefficients $\rhok$ used for each constraint were optimized by hand, through trial and error. Also, we note that when applying our method to the 2-D $(1,\infty)$-RLL constraint, our bound was inferior to the one presented in \cite{WeeksBlahut:98} (utilizing the method of \cite{CalkinWilf:97}).

\begin{table*}
\caption{Upper-bounds on the capacity of some 2-D constraints.}
\label{tbl:ourUpperBounds}
\begin{center}
\begin{tabular}{||c|c|c|c|c|l|l|l||}
\hline
Constraint & $r$ & $s$ & $k$ & $\prec$ used & Upper bound & Comparison & Lower bound \\
\hline \hline
$(2,\infty)$-RLL & 3 & 8 & 7 & $\lexprec$, $\skipprec$ &\; \; 0.4457 & 0.4459 \quad \cite{ForchhammerLaursen:07} & 0.444202 \cite{TalRoth:08} \\
\hline
$(3,\infty)$-RLL & 4 & 8 & 5 & $\lexprec$ & \; \; 0.36821 & 0.3686 \quad \cite{ForchhammerLaursen:07} & 0.365623 \cite{SharovRoth:08}\\
\hline
$(0,2)$-RLL & 3 & 5 & 2 & $\lexprec$ & \; \; 0.816731 & 0.817053 & 0.816007 \cite{SharovRoth:08}\\
\hline
n.i.b. & 3 & 4 & 1 & $\skipprec$ &  \; \; 0.92472 & 0.927855 & 0.922640 \cite{TalRoth:08} \\
\hline
\end{tabular}
\end{center}
\end{table*}

\section{Asymptotic analysis}

\label{sec:asymptotics}
For a given constraint $\calS$ and positive integers $r$ and $s$, let $t$ be an integer such that
$0 \leq t < s$. Denote by $\mu(r,s,t)$ the value of the
optimization program in Figure~\ref{fig:optProgram}, where the parameters are as in (\ref{eq:simplestCase}). In this section, we show that even if we restrict ourselves to this simple case, we get an upper bound which is asymptotically tight, in the following sense.

\begin{theo}
\label{theo:asymptoticallyTight}
For all $\epsilon > 0$, there exist
\[
r_0 > 0 \;, \quad s_0 > 0 \; , \quad 0 \leq t_0 < s_0 \;
\]
such that for all
\[
r \geq r_0 \; ,  \quad s \geq s_0 \; , \quad t_0 \leq t \leq s - (s_0-t_0) \; ,
\]
we have that
\[
\mu(r,s,t) - \capacity(\calS)  \leq \epsilon\; .
\]
\end{theo}

In order to prove Theorem~\ref{theo:asymptoticallyTight}, we need the following
lemma.

\begin{lemm}
\label{lemm:asymptoticallyTight}
For all $\epsilon > 0$, there exist
\[
r_0 > 0 \;, \quad s_0 > 0 \; , \quad 0 \leq t_0 < s_0 \;
\]
such that
\[
\mu(r_0,s_0,t_0) - \capacity(\calS)  < \epsilon\; .
\]
\end{lemm}
\begin{proof}
Another well known method for bounding $\capacity(\calS)$ from above is the so called
``stripe method'', mentioned in the introduction. Namely, for some given $\striplength$, consider the 1-D constraint $S=S(\striplength)$
defined as follows. The alphabet of the constraint is $\Sigma^{\striplength}$. A word of
length $r'$ satisfies $S$ if and only if when we write its entries
as rows of length $\striplength$, one below the other, we get an $r' \times \striplength$ configuration which
satisfies the 2-D constraint $\calS$.

Define the normalized capacity of $S$ as
\[
\ncapacity(S) = \frac{1}{\striplength} \capacity(S) \; .
\]
By the definition of $\capacity(\calS)$, the normalized capacity
approaches $\capacity(\calS)$ as $\striplength \to \infty$. Thus, fix a $\striplength$ such that
\[
\ncapacity(S) - \capacity(\calS) \leq \epsilon/2 \; .
\]

We say that a 1-D constraint has memory $\sfm$ if there exists a graph representing it, and all paths in the graph of length $\sfm$ with the same labeling terminate in the same vertex.
By \cite[Theorem 3.17]{MarcusRothSiegel:98} and its proof, there exists a series of 1-D constraints $\set{S_\sfm}_{\sfm=1}^\infty$ such that $S \subseteq S_\sfm$, the memory of $S_\sfm$ is $\sfm$, and $\lim_{\sfm \to \infty} \capacity(S_\sfm) = \capacity(S)$. Thus, fix $\sfm$ such that
\[
\ncapacity(S_\sfm) - \ncapacity(S) \leq \epsilon/2 \; .
\]

To finish the proof, we now show that
\[
\mu(r_0,s_0,t_0) \leq \ncapacity(S_\sfm)  \; ,
\]
where
\[
r_0=\sfm+1 \; , \quad s_0 = 2 \cdot \striplength \; , \quad t_0=\striplength-1 \; .
\]
Note that $\mu(r_0,s_0,t_0)$ is the maximum of
\begin{equation}
\label{eq:initialAsymptotic}
H(\bar{X}_{\sfm,\striplength-1} | \bar{X}[\Tis_{\sfm,\striplength-1}^{(\lexprec)} \cap \Bis_{\sfm+1,2 \cdot \striplength}]) \; ,
\end{equation}
over all random variables $\bar{X} \in \calS_{\sfm+1,2 \cdot \striplength}$ with a
probability distribution satisfying our linear requirements.

For all
$0 \leq \phi < \striplength$ we get by the (imposed) stationarity of $\bar{X}$ that
(\ref{eq:initialAsymptotic}) is bounded from above by
\[
H_\phi = H(\bar{X}_{\sfm,\phi} | \bar{X}[\Tis_{\sfm,\phi}^{(\lexprec)} \cap \Bis_{\sfm+1,\striplength}] )
\; .
\]
So, (\ref{eq:initialAsymptotic}) is also bounded from above by
\begin{equation}
\label{eq:asymptoticsSumAlpha}
\frac{1}{\striplength} \sum_{\phi =0 }^{\striplength-1} H_\phi \; .
\end{equation}
The first $\striplength$ columns of $\bar{X}$ form a configuration with index set $\Bis_{\sfm+1,\striplength}$. By our linear
requirements, stationarity (specifically, vertical stationarity) holds for this
configuration as well. So, we may define a stationary 1-D Markov chain \cite[\S 3.2.3]{MarcusRothSiegel:98} on $S_\sfm$, with
entropy given by (\ref{eq:asymptoticsSumAlpha}). That entropy, in turn, is at most $\ncapacity(S_\sfm)$.
\end{proof}

\begin{proof}[Proof of Theorem \ref{theo:asymptoticallyTight}]
The following inequalities are easily verified:
\[
\mu(r,s,t) \geq \mu(r+1,s,t) \; .
\]
\[
\mu(r,s,t) \geq \mu(r,s+1,t) \; .
\]
\[
\mu(r,s,t) \geq \mu(r,s+1,t+1) \; .
\]
The proof follows from them and Lemma~\ref{lemm:asymptoticallyTight}.
\end{proof}


\section*{Appendix}
Our goal in this appendix is to prove Theorem~\ref{theo:burtonSteif}. Essentially, Theorem~\ref{theo:burtonSteif} will turn out to be a corollary of \cite[Theorem 1.4]{BurtonSteif:94}. However, \cite[Theorem 1.4]{BurtonSteif:94} deals with configurations in which the index set is $\integers^2$. So, some definitions and auxiliary lemmas are in order.

Recall that $(\Gr,\Gc)$ is the pair of vertex-labeled graphs through which $\calS=\calS(\Gr,\Gc)$ is defined. Also, recall that each member of $\calS$ is a configuration with a rectangular index set. Namely, the index set of a configuration in $\calS$ is $\sigma_{i,j}(\Bis_{M,N})$, for some $i$, $j$, $M$, and $N$. We now give a very similar definition to that of $\calS$, only now we require that the index set of each configuration is $\integers^2$. Namely, define $\scriptS=\scriptS(\Gr,\Gc)$ as follows: A configuration $(w_{i,j})_{(i,j) \in \integers^2}$ over $\Sigma$ is in $\scriptS(\Gr,\Gc)$ iff there exists a configuration $(u_{i,j})_{(i,j) \in \integers^2}$ over the vertex set $V$ with the following properties: for all $(i,j) \in \integers^2$, (a) the labeling of $u_{i,j}$ satisfies $L(u_{i,j})= w_{i,j}$; (b) there exists an edge from $u_{i,j}$ to $u_{i,j+1}$ in $\Gr$; (c) there exists an edge from $u_{i,j}$ to $u_{i+1,j}$ in $\Gc$.

For positive integers $M,N > 0$, define $\scriptS_{M,N}$ as the restriction of $\scriptS$ to $\Bis_{M,N}$. Namely,
\[
\scriptS_{M,N} = \scriptS[\Bis_{M,N}] \; ,
\]
where the definition of the restriction operation is as in (\ref{eq:restriction}).
Also, for $M$ equal to $N$, define
\[
\scriptS_{M} = \scriptS_{M,M} \; .
\]
Note that for all $M,N > 0$ we have
\begin{equation}
\label{eq:scriptInCal}
\scriptS_{M,N} \subseteq \calS_{M,N} \; ,
\end{equation}
and there are cases in which the inclusion is strict. Next, define the capacity of $\scriptS$ as
\[
\capacity(\scriptS) = \lim_{M \to \infty}
\frac{1}{M^2} \cdot \log_2 \size{\scriptS_M} \; .
\]
The limit indeed exists, by sub-additivity (see \cite[Appendix]{KatoZeger:99}, and references therein).

For integers $M,N>0$ and $\delta \geq 0$, denote
\[
\Cis_{M,N,\delta} = \sigma_{-\delta,-\delta}(\Bis_{M+2\delta,N+2\delta})
\]
and let
\[
\calS_{M,N,\delta} = \calS[\Cis_{M,N,\delta}] \; .
\]
Note that the index set $\Cis_{M,N,\delta}$ of each element of $\calS_{M,N,\delta}$ is simply $\Bis_{M,N}$, padded with $\delta$ columns to the right and left and $\delta$ rows to the top and bottom.
The following lemma will help us bridge the gap between finite and infinite index sets.
\begin{lemm}
\label{lemm:finiteToInfinate}
Let $w$ be a configuration over the finite alphabet $\Sigma$ with index set $\Bis_{M,N}$. If for all $\delta \geq 0$ we have that
\begin{equation}
\label{eq:wCanBeEnlarged}
w \in \calS_{M,N,\delta}[\Bis_{M,N}] \; ,
\end{equation}
then we must have that
\[
w \in \scriptS_{M,N} \; .
\]
\end{lemm}

\begin{proof}
Define the following auxiliary directed graph. The vertex set is
\[
\bigcup_{\delta \geq 0} \set{\hat{w} \in \calS_{M,N,\delta} \; : \; \hat{w}[\Bis_{M,N}] = w} \; .
\]
For every $\delta \geq 0$, there is a directed edge from $w_1 \in \calS_{M,N,\delta}$ to $w_2 \in \calS_{M,N,\delta+1}$ iff $w_1 = w_2[\Cis_{M,N,\delta}]$. It is easily seen that this graph is a directed tree with root $w$, as defined in \cite[\S 2.4]{Even:79}. Since (\ref{eq:wCanBeEnlarged}) holds for all $\delta \geq 0$, the vertex set of the tree is infinite (and countable). On the other hand, since the alphabet size $\size{\Sigma}$ is finite, the out-degree of each vertex is finite. Thus, by K\"{o}nig's Infinity Lemma \cite[Theorem 2.8]{Even:79}, we must have an infinite path in the tree starting from the root $w$.

Denote the vertices of the above-mentioned infinite path as
\[
w = w^{[0]},w^{[1]},w^{[2]},\ldots \; .
\]
We now show how to find a configuration $(w'_{i,j})_{(i,j) \in \integers^2}$ such that $w' \in \scriptS$ and $w = w'[\Bis_{M,N}]$. For each $(i,j) \in \integers^2$, define $w'_{i,j}$ as follows: let $\delta \geq 0$ be such that $(i,j) \in \Cis_{M,N,\delta}$, and take $w'_{i,j} = w^{[\delta]}_{i,j}$. It is easily seen that $w'$ is well defined and contained in $\scriptS$.
\end{proof}

The following lemma states that although the inclusion in (\ref{eq:scriptInCal}) may be strict, the capacities of $\calS$ and $\scriptS$ are equal.
\begin{lemm}
Let $\calS$ and $\scriptS$ be as previously defined. Then,
\begin{equation}
\label{eq:capacitiesEqual}
\capacity(\scriptS) = \capacity(\calS) \; .
\end{equation}
\end{lemm}

\begin{proof}
By (\ref{eq:scriptInCal}), we must have that $\capacity(\scriptS) \leq \capacity(\calS)$. For the other direction, it suffices to prove that for all $M > 0$,
\begin{equation}
\label{eq:capacityUpperBoundedByInfiniteSets}
\capacity(\calS) \leq \frac{1}{M^2} \log_2 \size{\scriptS_M} \; .
\end{equation}
So, let us fix $M$ and prove the above. By Lemma~\ref{lemm:finiteToInfinate}, there exists $\delta \geq 0$ such that for all $w \in \Sigma^{M \times M}$,
\[
w \not\in \scriptS_M \quad \Longrightarrow \quad w \not\in \calS_{M,M,\delta}[\Bis_{M}] \; .
\]
For $t >0$, let $M'$ be shorthand for
\[
M' = t \cdot M \; .
\]
By the definition of capacity, we have that
\begin{equation}
\label{eq:capacityAsJumps}
\capacity(\calS) = \lim_{t \to \infty} \frac{1}{(M')^2} \log_2 \size{\calS_{M'}} \; .
\end{equation}
Now, let us partition $\Bis_{M'}$ into the following disjoint sub-sets of indexes: for $0 \leq i,j < t$, define the set
\[
\Dis_{i,j} = \sigma_{i \cdot M, j \cdot M}(\Bis_{M}) \; .
\]
Let $w' \in \calS_{M'}$. Notice that for all $0 \leq i,j < t$ for which
\begin{equation}
\label{eq:DisInside}
\sigma_{i \cdot M, j \cdot M}(\Cis_{M,M,\delta}) \subseteq \Bis_{M'} \; ,
\end{equation}
we must have that $w'[\Dis_{i,j}]$ is equal to some correspondingly shifted element of $\scriptS_M$. On the other hand, for $M$ and $\delta$ fixed, the number of pairs $(i,j)$ for which (\ref{eq:DisInside}) does not hold is $O(t)$. Thus, a simple calculation gives us that
\[
\frac{1}{(M')^2} \log_2 \size{\calS_{M'}} \leq \frac{1}{M^2} \log_2 \size{\scriptS_{M}} + O(1/t) \; .
\]
This, together with (\ref{eq:capacityAsJumps}), proves (\ref{eq:capacityUpperBoundedByInfiniteSets}).
\end{proof}

For a given $M > 0$, define the set $\scriptF(M)$ of configurations with index set $\integers^2$ as follows: a configuration $(w_{i,j})_{(i,j)\in \integers^2}$ is in $\scriptF(M)$ iff for all $(i,j) \in \integers^2$,
\[
w[\sigma_{i,j}(\Bis_M)] \in \scriptS_M \; .
\]
Namely, each $M \times M$ ``patch'' is a correspondingly shifted element of $\scriptS_M$.

Note that there exist vertex-labeled graphs $\Gr(M)$ and $\Gc(M)$ such that $\scriptF(M) = \scriptS(\Gr(M),\Gc(M))$. Specifically, the vertex set of both graphs is equal to $\scriptS_M$; the label of each such vertex is its lower-left entry; there is an edge from $w_1 \in \scriptS_M$ to $w_2 \in \scriptS_M$ in $\Gr(M)$ ($\Gc(M)$) iff the first $M-1$ rows (columns) of $w_1$ are equal to the last $M-1$ (rows) columns of $w_2$. Thus, $\capacity(\scriptF(M))$ exists. Also, since $w \in \scriptS$ implies $w \in \scriptF(M)$, we have
\begin{equation}
\label{eq:capacityOfFMLowerBounded}
\capacity(\scriptS) \leq \capacity(\scriptF(M)) \; .
\end{equation}
The following is a direct corollary of \cite[Theorem 1.4]{BurtonSteif:94}.
\begin{coro}
\label{coro:FM}
For all $M > 0$, there exists a stationary random variable $W^{(M)}$ taking values on $\scriptF(M)[\Bis_M]$ such that
\begin{equation}
\label{eq:capacityOfFMUpperBounded}
\capacity(\scriptF(M)) \leq \frac{1}{M^2} H(W^{(M)}) \; .
\end{equation}
\end{coro}

\begin{proof}[Proof of Theorem~\ref{theo:burtonSteif}]
Notice that
\[
\scriptF(M)[\Bis_M] = \scriptS_M \subseteq \calS_M \; .
\]
Thus, take $W^{(M)}$ as in Corollary~\ref{coro:FM} and notice that it satisfies conditions (i) and (ii) in Theorem~\ref{theo:burtonSteif}. From (\ref{eq:capacitiesEqual}), (\ref{eq:capacityOfFMLowerBounded}), and (\ref{eq:capacityOfFMUpperBounded}) we get that
\[
\capacity(\calS) \leq \lim_{M \to \infty}
\frac{1}{M^2} \cdot H(W^{(M)}) \; .
\]
But since $W^{(M)}$ takes values on $\calS_M$, we have by \cite[Page 19]{CoverThomas:91} that the above inequality is in fact an equality. Thus, condition (iii) is proved.
\end{proof}
%
%
\twobibs{
\bibliographystyle{IEEEtran}
\bibliography{../../mybib}

\begin{thebibliography}{10}

\bibitem{MarcusRothSiegel:98}
B.~H. Marcus, R.~M. Roth, and P.~H. Siegel, ``Constrained systems and coding
  for recording channels,'' in \emph{Handbook of Coding Theory}, V.~Pless and
  W.~Huffman, Eds.\hskip 1em plus 0.5em minus 0.4em\relax Amsterdam: Elsevier,
  1998, pp. 1635--1764.

\bibitem{WeeksBlahut:98}
W.~Weeks and R.~E. Blahut, ``The capacity and coding gain of certain
  checkerboard codes,'' \emph{IEEE Trans. Inform. Theory}, vol.~44, pp.
  1193--1203, 1998.

\bibitem{KatoZeger:99}
A.~Kato and K.~Zeger, ``On the capacity of two-dimensional run-length
  constrained code,'' \emph{IEEE Trans. Inform. Theory}, vol.~45, pp.
  1527--1540, 1999.

\bibitem{Etzion:97}
T.~Etzion, ``Cascading methods for runlength-limited arrays,'' \emph{IEEE
  Trans. Inform. Theory}, vol.~43, pp. 319--324, 1997.

\bibitem{Halevy+:04}
S.~Halevy, J.~Chen, R.~M. Roth, P.~H. Siegel, and J.~K. Wolf, ``Improved
  bit-stuffing bounds on two-dimensional constraints,'' \emph{IEEE Trans.
  Inform. Theory}, vol.~50, pp. 824--838, 2004.

\bibitem{ForchhammerJustesen:00}
S.~Forchhammer and J.~Justesen, ``Bounds on the capacity of constrained
  two-dimensional codes,'' \emph{IEEE Trans. Inform. Theory}, vol.~46, pp.
  2659--2666, 2000.

\bibitem{CalkinWilf:97}
N.~Calkin and H.~S. Wilf, ``The number of independent sets in a grid graph,''
  \emph{SIAM J. Discrete Math.}, vol.~11, pp. 54--60, 1997.

\bibitem{BurtonSteif:94}
R.~Burton and J.~E. Steif, ``Non-uniqueness of measures of maximal entropy for
  subshifts of finite type,'' \emph{Ergod. Th. Dynam. Sys.}, vol.~14, pp.
  213--235, 1994.

\bibitem{CoverThomas:91}
T.~M. Cover and J.~A. Thomas, \emph{Elements of Information Theory}.\hskip 1em
  plus 0.5em minus 0.4em\relax Wiley, 1991.

\bibitem{BoydVandenberghe:04}
S.~Boyd and L.~Vandenberghe, \emph{Convex Optimization}.\hskip 1em plus 0.5em
  minus 0.4em\relax Cambridge, UK: Cambridge University Press, 2004.

\bibitem{Czyzyk+:98}
J.~Czyzyk, M.~P. Mesnier, and J.~J. Mor\'{e}, ``The {NEOS} server,'' \emph{IEEE
  Computational Science~\& Engineering}, vol.~5, no.~3, pp. 68--75, 1998.

\bibitem{GroppMore:97}
W.~Gropp and J.~J. Mor\'{e}, ``Optimization environments and the {NEOS}
  server,'' in \emph{Approximation Theory and Optimization}.\hskip 1em plus
  0.5em minus 0.4em\relax Cambridge University Press, 1997, pp. 167--182.

\bibitem{Dolan+:02}
E.~D. Dolan, R.~Fourer, J.~J. Mor\'{e}, and T.~S. Munson, ``The {NEOS} server
  for optimization: Version 4 and beyond,'' Mathematics and Computer Science
  Division, Argonne National Laboratory, Argonne, IL, Tech. Rep., 2002.

\bibitem{Fourer+:02}
R.~Fourer, D.~M. Gay, and B.~W. Kernighan, \emph{AMPL: A Modeling Language for
  Mathematical Programming}, 2nd~ed.\hskip 1em plus 0.5em minus 0.4em\relax
  Duxbury Press, 2002.

\bibitem{WachterBiegle:06}
A.~W{\"a}chter and L.~T. Biegler, ``On the implementation of an interior-point
  filter line-search algorithm for large-scale nonlinear programming,''
  \emph{Math. Program.}, vol. 106, no.~1, pp. 25--57, 2006.

\bibitem{ForchhammerLaursen:07}
S.~Forchhammer and T.~V. Laursen, ``Entropy of bit-stuffing-induced measures
  for two-dimensional checkerboard constraints,'' \emph{IEEE Trans. Inform.
  Theory}, vol.~53, pp. 1537--1546, 2007.

\bibitem{TalRoth:08}
I.~Tal and R.~M. Roth, ``Bounds on the rate of 2-{D} bit-stuffing encoders,''
  in \emph{Proc. IEEE Int'l Symp. Inform. Theory (ISIT'2008)}, Toronto,
  Ontario, Canada, 2008.

\bibitem{SharovRoth:08}
A.~Sharov and R.~M. Roth, ``Two-dimensional constrained coding based on
  tiling,'' in \emph{Proc. IEEE Int'l Symp. Inform. Theory (ISIT'2008)},
  Toronto, Ontario, 2008, pp. 1468--1472.

\bibitem{Even:79}
S.~Even, \emph{Graph Algorithms}.\hskip 1em plus 0.5em minus 0.4em\relax
  Computer Science Press, 1979.

\end{thebibliography}
}
{

}
\end{document}